# US HL-LHC Accelerator Upgrade Project
# MQXFA11 shipping incident report

M. Baldini[1], G. Ambrosio[1], J. Blower[1], J. Cozzolino[2], T. Strauss[1], G. Vallone[3]

1Fermi National Accelerator Laboratory, Batavia, IL 60510 USA
2Brookhaven National Laboratory, Upton, NY 11973 USA
3Lawrence Berkeley National Laboratory, Berkeley, CA 94720 USA.

## TABLE OF CONTENTS

## 1. Introduction

The truck transporting the MQXFA11 magnet from LBNL to BNL was involved in an accident on 7/20/22. It was rear ended by another truck. Drivers were not injured.

The main hit took place on the right back corner (see Fig. 1). During the incident the truck rear axle as displayed in Fig. 1 (right panel).

The truck with the MQXFA11 magnet was towed to a tow yard in Rockdale, IL (~50 min drive from Fermilab). Subsequently, it was moved on another truck and moved to Fermilab. Fermilab employees added 3 accelerometers on the crate before it was moved on the new truck and attended all loading/unloading operations at the tow yard and at FNAL.

The magnet arrived at FNAL on July 28th. Upon arrival a visual inspection was performed. Accelerometers were removed and shock data were analyzed. Electrical checkout and a metrology survey were carried out. Strain data were also collected from the fiber optics installed on two coils and compared with data obtained before shipment.

This report presents a summary of results. More pictures and details can be found at: https://indico.fnal.gov/event/55698/

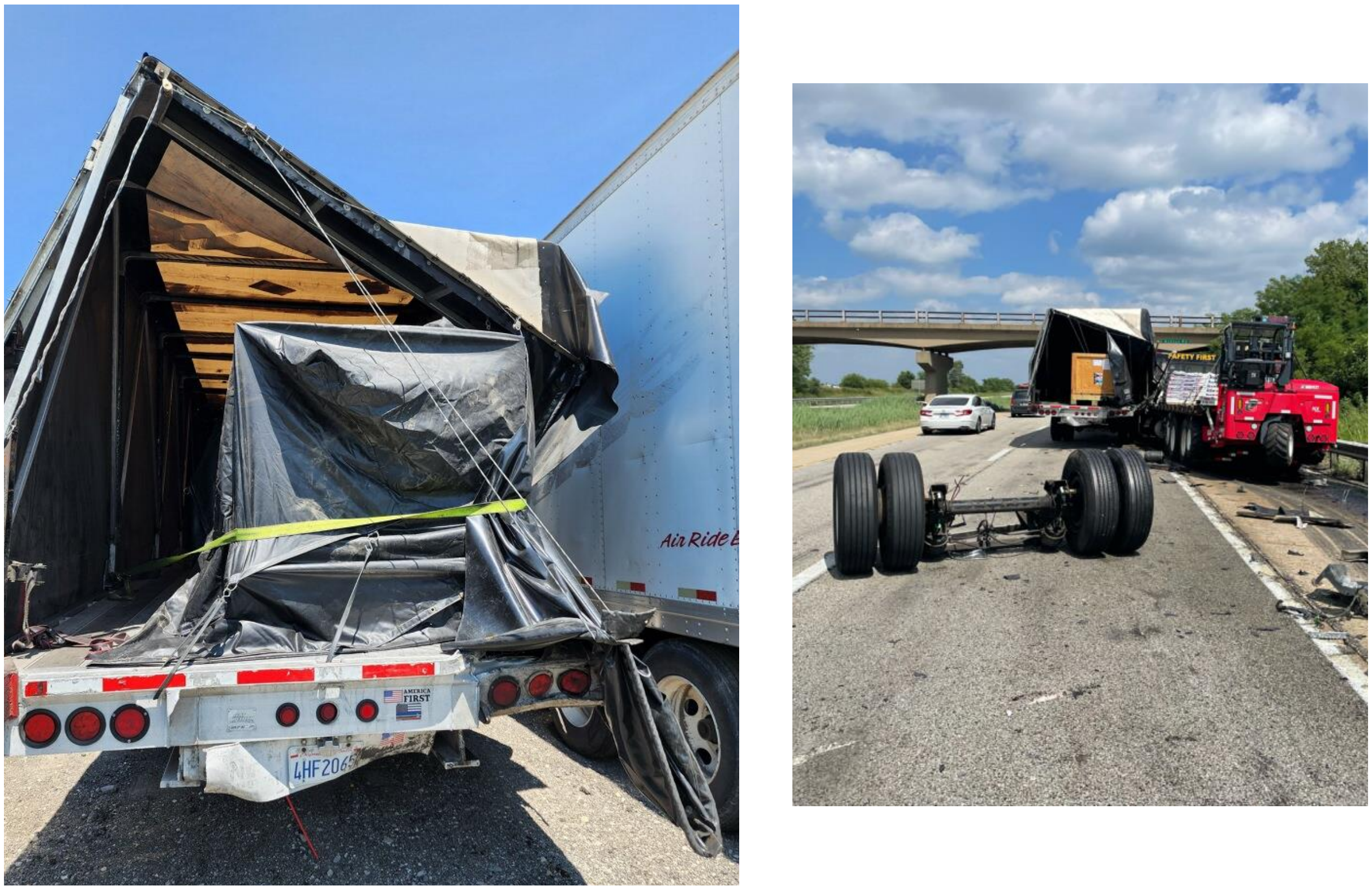

Figure 1: Pictures of the truck taken right after the incident.

## 2. Visual inspection at the tow yard

The center of the crate was sitting 17’4’’ away from the rear end of the trailer. The lead end of the magnet is on the cab end of the trailer (see Fig. 1).

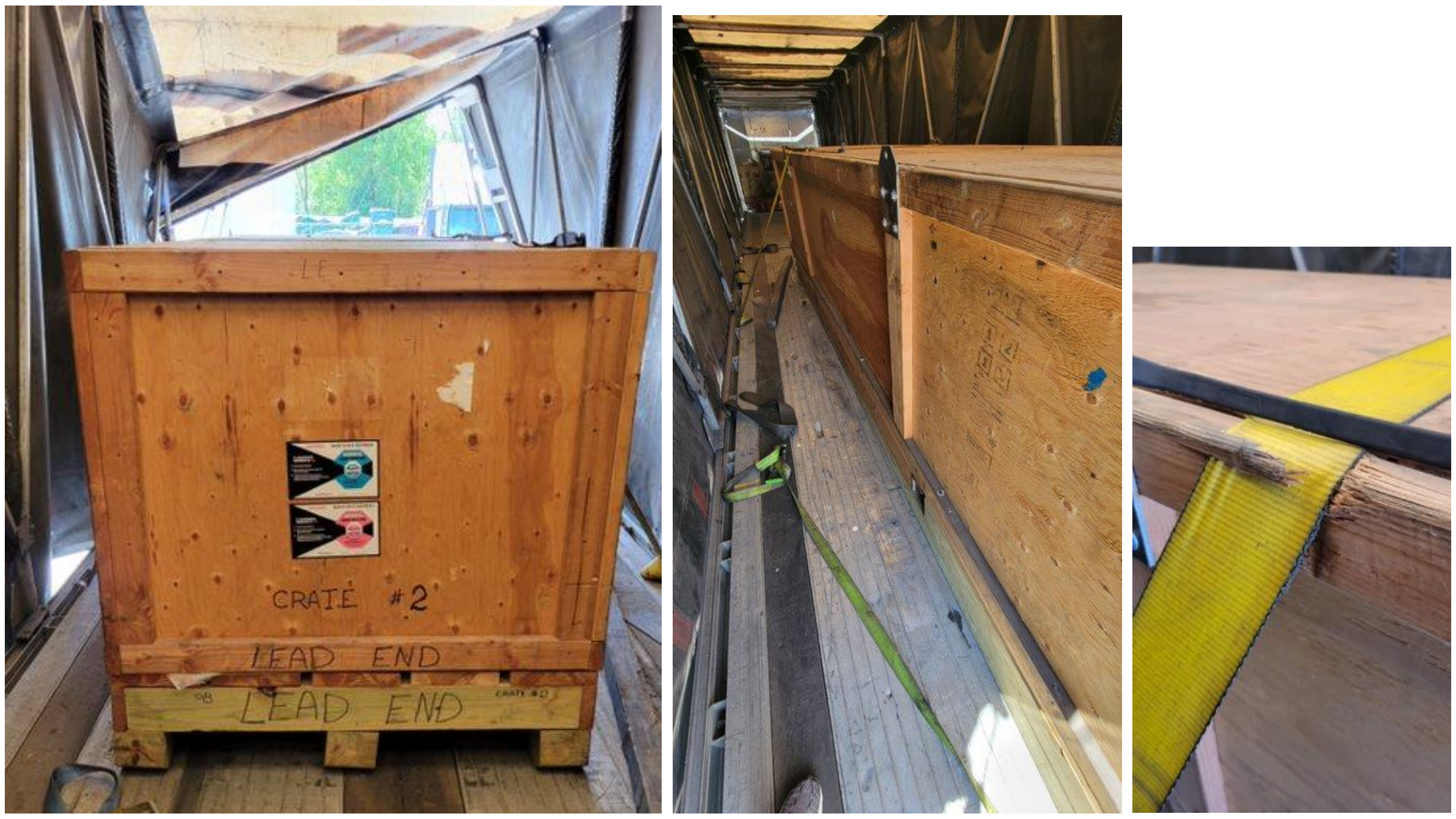

Figure 1: Left to right: Photo of the lead end of the crate/magnet taken after the incident from the cab end of the trailer; The yellow sling was found close to the metallic tab; detail of the yellow ling after the incident

The magnet was also tied down to the truck using two lings attached to two tabs.

The rear end of the crate was found shifted toward the passenger side of the trailer. One of the slings was broken and the crate moved backward 6-7’ according to the position of the undamaged sling (see details in Fig. 1).

Two shock watches vertically oriented were installed on the lead end driver side and on the non-lead end passenger side of the magnet crate.

Both 5g and 10 g shock watches installed on non-lead end passenger side tripped.

### 3. Inspection at FNAL upon arrival

The magnet was found in very good conditions. In Fig.2 pictures of the end plate, the bore and the longitudinal stopper are displayed. There are no signs of damage.

The arrows define the vertical, the transverse and the axial directions.

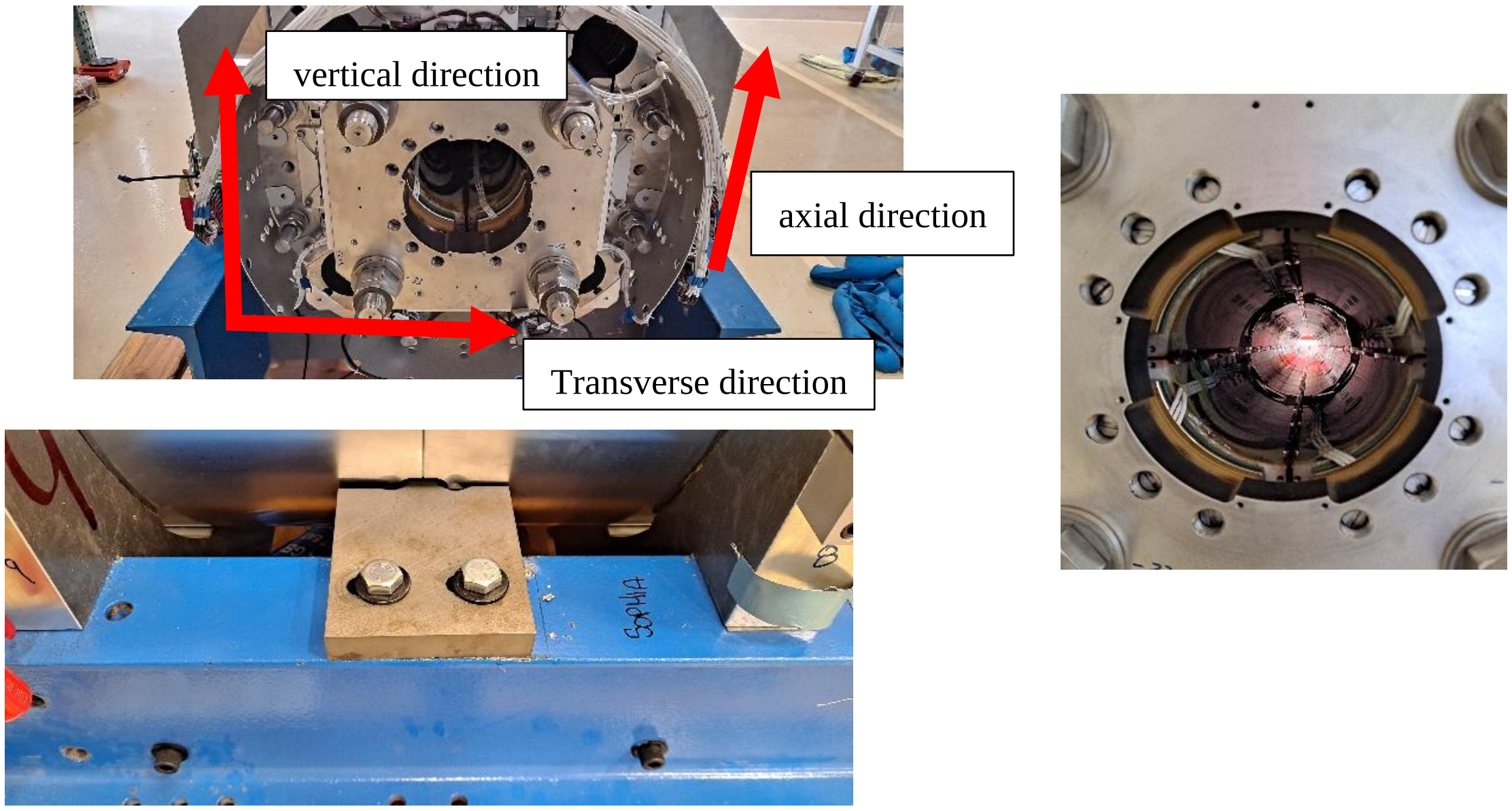

Figure 2: Photos of the return end, magnet bore, and longitudinal stopper taken at FNAL during the visual investigation

Several high voltage tests were performed. In Table. 1 a summary of the results is shown. Coil to ground (GND), strip heater (SH) to coil hipot were all successful

| Test Item | Target HV | Dwell Time | Voltage Lead Location | Ground Location | Success? |
|---|---|---|---|---|---|
| Coil - GND | 3680 | 30 | Coil Lead | Magnet Shell | YES |
| SH - Coil | 3680 | 30 | SH Lead | Coil lead | YES |
| Coil - SH | 3680 | 30 | Coil Lead | SH lead | YES |
| SH - GND | 3680 | 30 | SH Lead | Magnet Shell | YES |

The gaps between the axial restrain plates and load keys were measured at the lead end and non-lead end of the magnet and compared with the measurements performed at LBNL before shipping. The results are in good agreement; the largest observed difference is around 300 um at the left top side of the lead end plate.

The distance from the yoke and the pizza box was measured together with the distance between the yoke and the rod end and compared with data taken before shipping. The results are in good agreement in within 0.5 mm (details can be found here: https://indico.fnal.gov/event/55698/).

Finally, a metrology survey was performed to check the MQXFA11 straightness. The measurements are taken longitudinally in seven points. It is worth noticing that the measurements were taken with the magnet sitting on a table at LBNL and with the magnet sitting on the shipping fixture at FNAL. X and Y values indicates a point on the magnet circumference along the transverse direction and the vertical direction as defined in Fig. 4. Data taken at LBNL and FNAL are show in Fig. 3. The general behavior of the data is in good agreement as well as the maximum deflection along y.

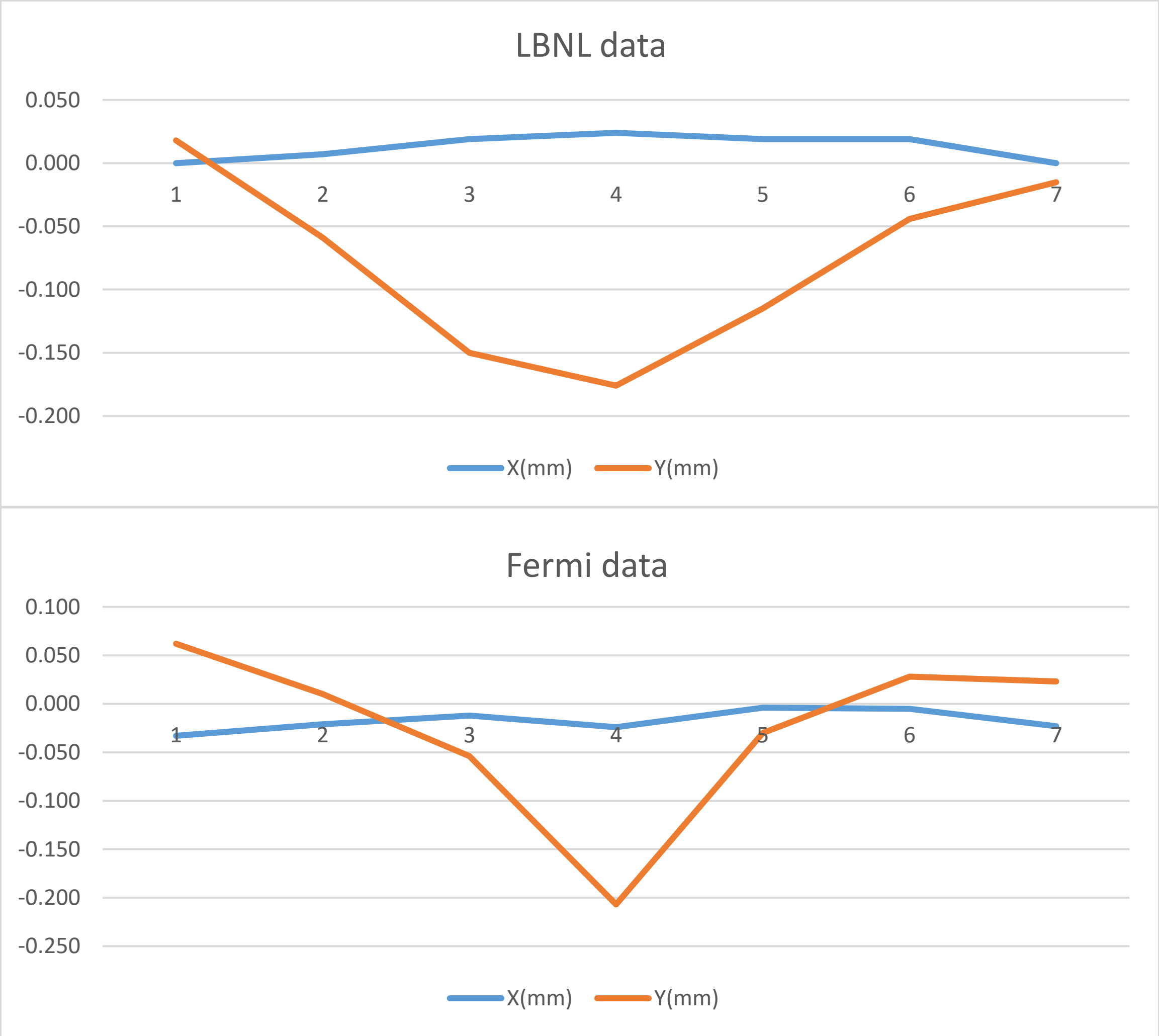

Figure 3: Measurements of magnet MQXFA11 straightness along the vertical and transverse directions taken at LBNL and FNAL.

## 4. Analysis of MQXFA11 accelerometer data during truck incident

In Fig. there is a picture of MQXFA11 after arriving at FNAL. Five accelerometers were installed on the magnet:

- Two GP1L devices were on top the fixture clamps: one on the lead end and one on the non-lead end.
- Three Slamstick devices were installed: one the end plate and on top of the lead end and on the side of the non-lead end fixture clamps (green circle).

The memory of the devices on the end plate and on lead end fixture clamp was full the day before the accident, therefore only three devices were collecting data during the accident: the two GP1L devices and the slamstick on the non-lead end clamp.

The axes of the accelerometers are oriented in a different way depending on the device position. To avoid any confusion, the reader should refer to the directions defined in Fig.4. Those directions are defined for both for the GP1L and the slamstick devices.

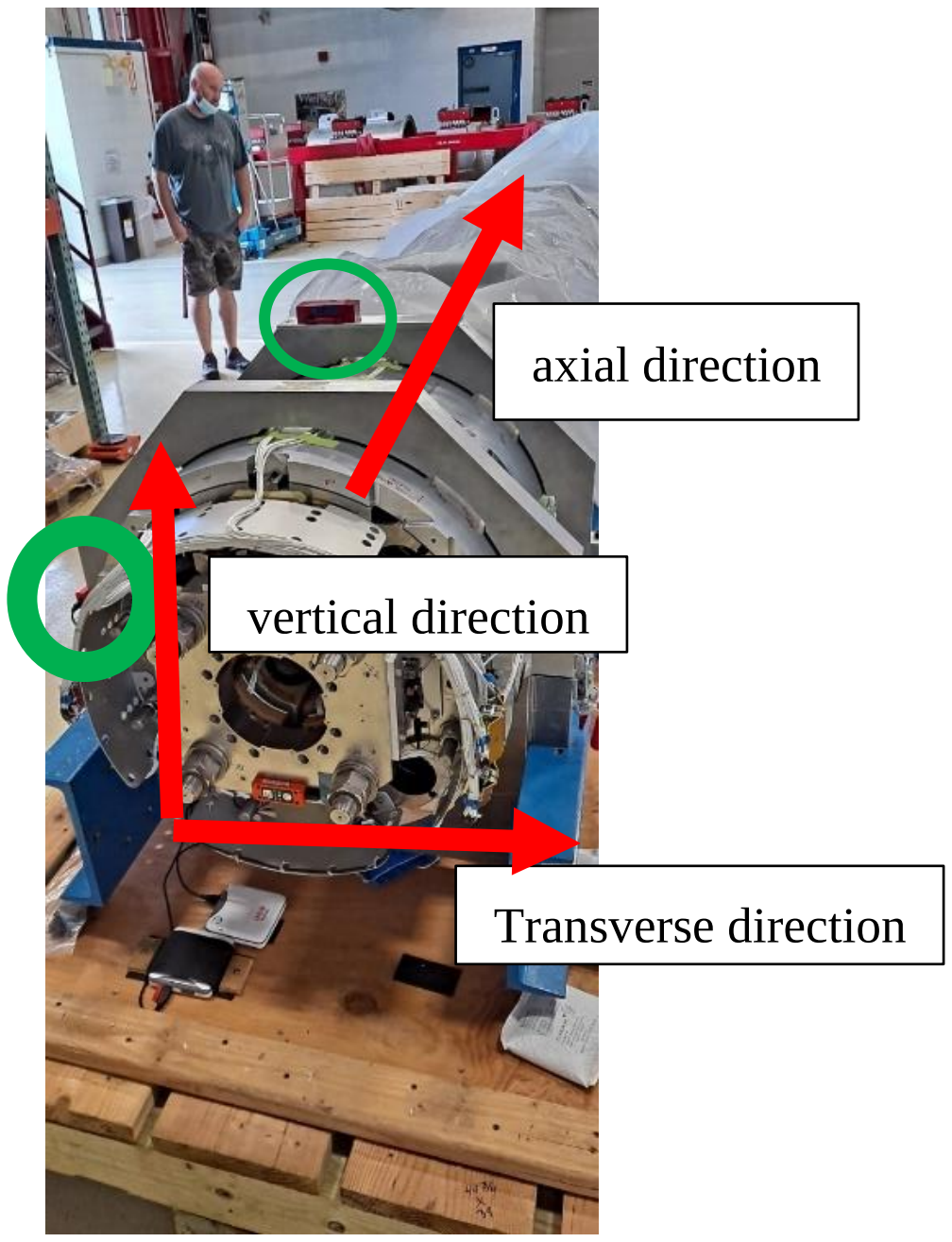

Figure 4: Picture of the MQXFA11 magnet. The green circles identify the accelerometer positions. The red arrows indicate the shock directions used in accelerometer data analysis

## 5. Shock observed with GP1L devices

The GP1L accelerometer settings are: 100 Hz sample rate and 50 Hz frequency response. the maximum measurable shock is 6.7g.

In Fig. 5 and 6, the lead end and the non-lead end shock snapshots of the entire trip is shown. The truck incident corresponds to the highest peaks observed in the two plots. it takes place on July 20$^{th}$ at 18:50 UTC. No data saturation is observed over the entire trip. The highest shocks (a little bit over 4g) are observed in the vertical and transverse direction in the non-lead end device. No vertical high shock is observed on the lead and device.

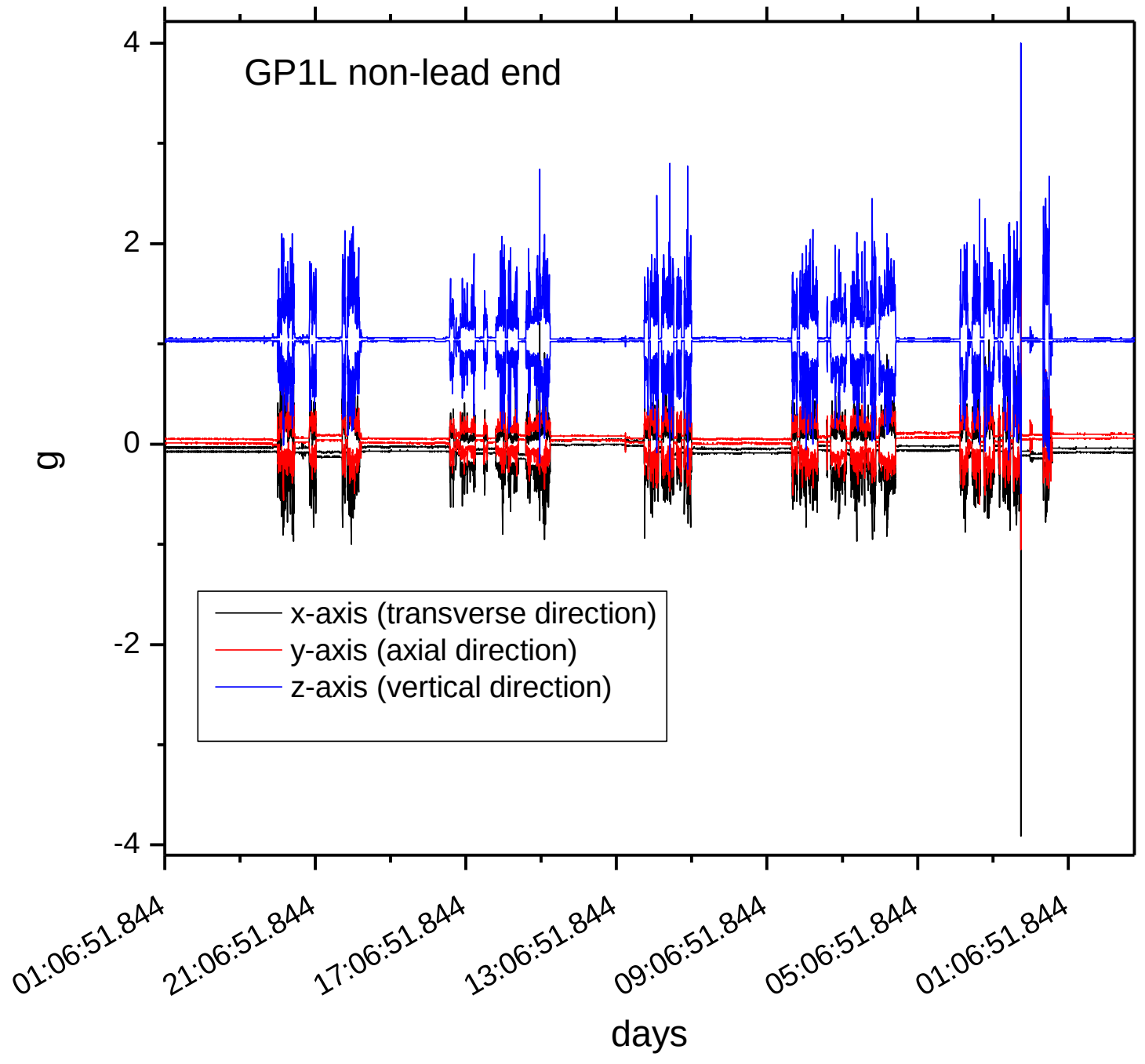

Figure 5: Snapshot of the shocks collected with the device installed on the non-lead end clamp

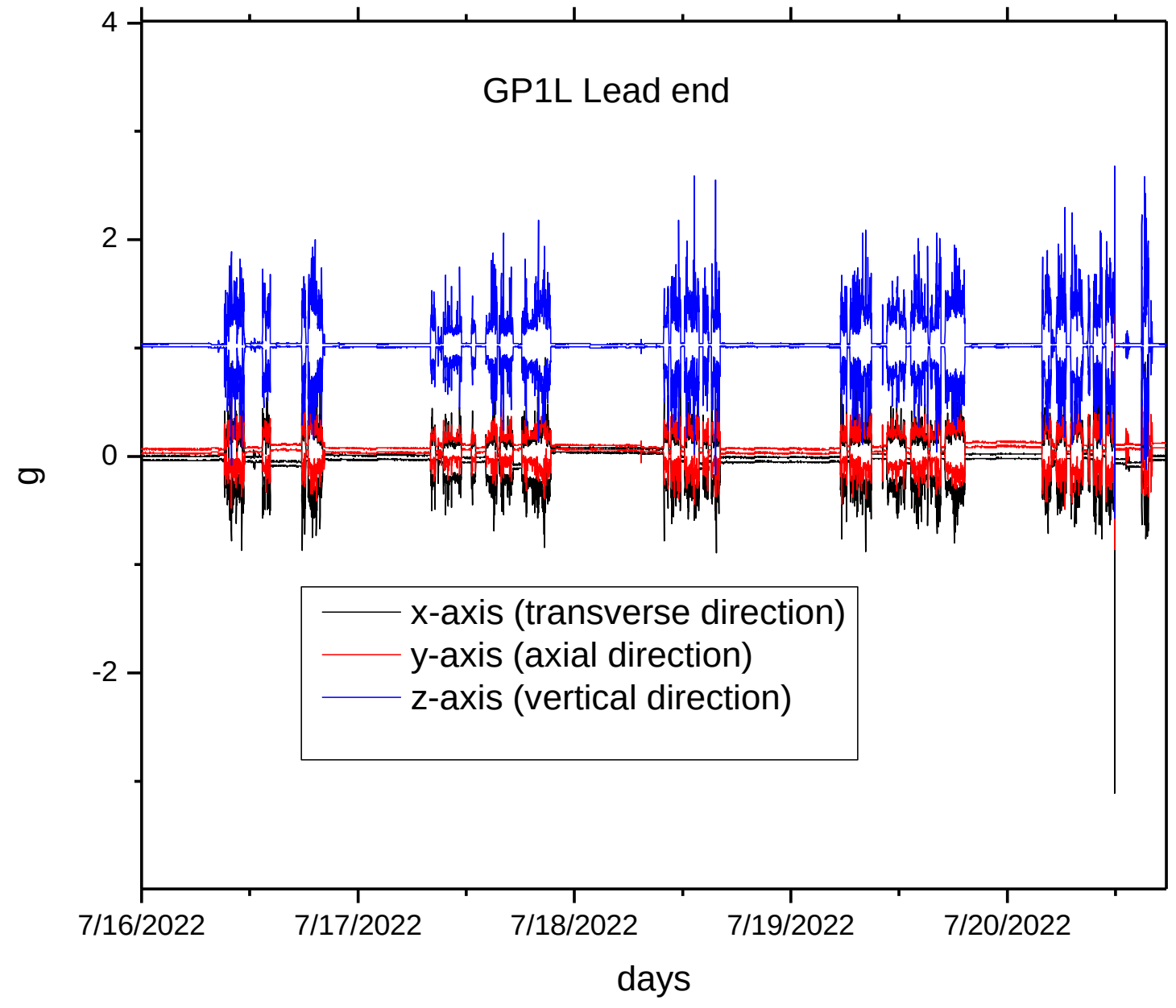

Figure 6: Snapshot of the shocks collected with the device installed on the lead end clamp

When g> 3g it is possible to record event with the GP1L device. The data are collected 0.3 s before and 0.7 s after the event. Each event is 1 sec long. Only two events above 3g were recorded during whole trip. Those events recorded the shocks during the accident. They occur one after the other in a two second timeframe. The two events are plotted together for each device (lead end and no- lead end) in Fig. 7 and Fig. 8. No shocks above 4g are recorded. Highest shocks are along the transverse and vertical direction which is consistent with the fact that the non-lead end of the crate moved a few inches away from the driver side

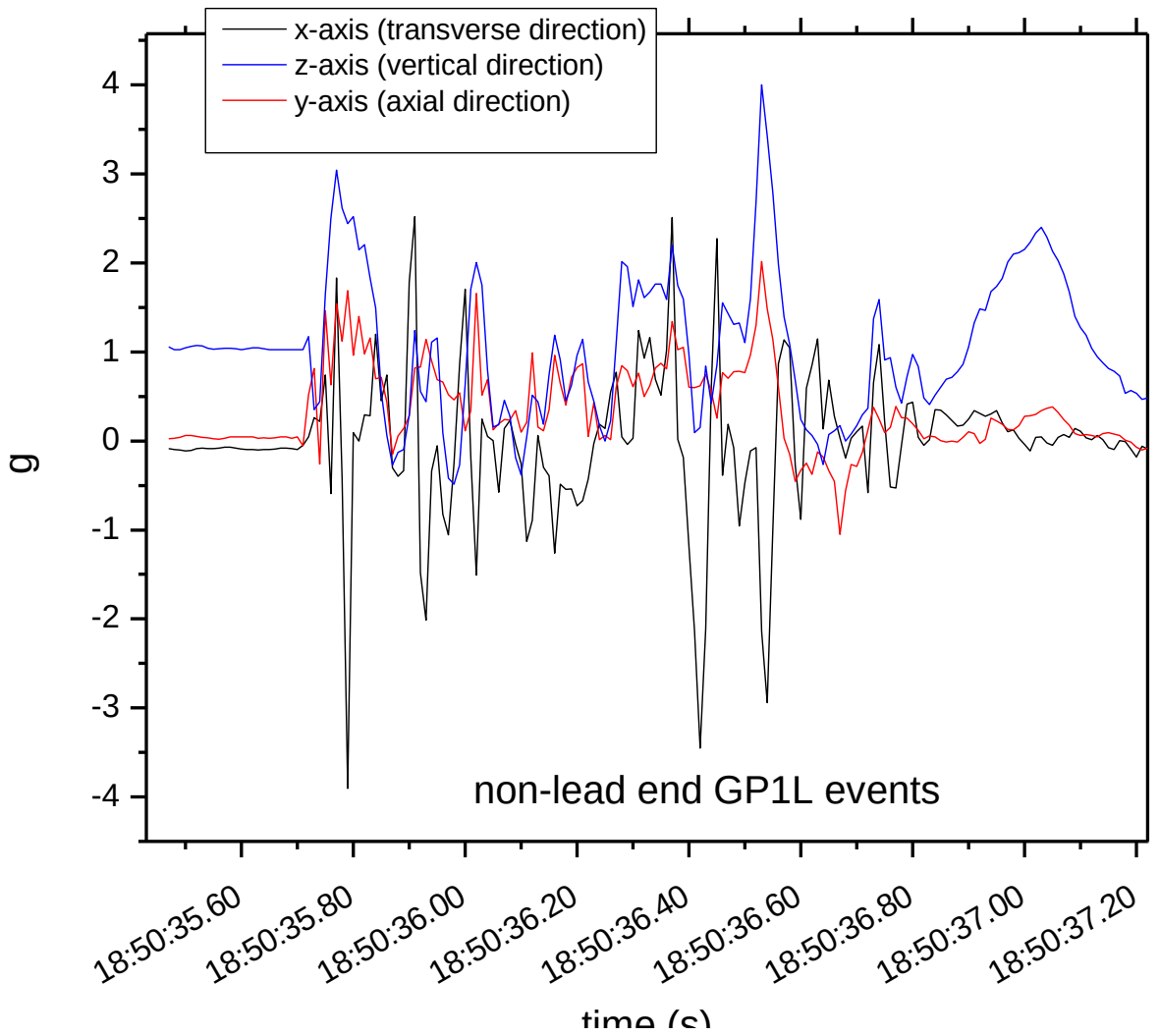

Figure 7: Two second event recorded by the non-lead end device. Event data are recorded when shocks are above 3 g.

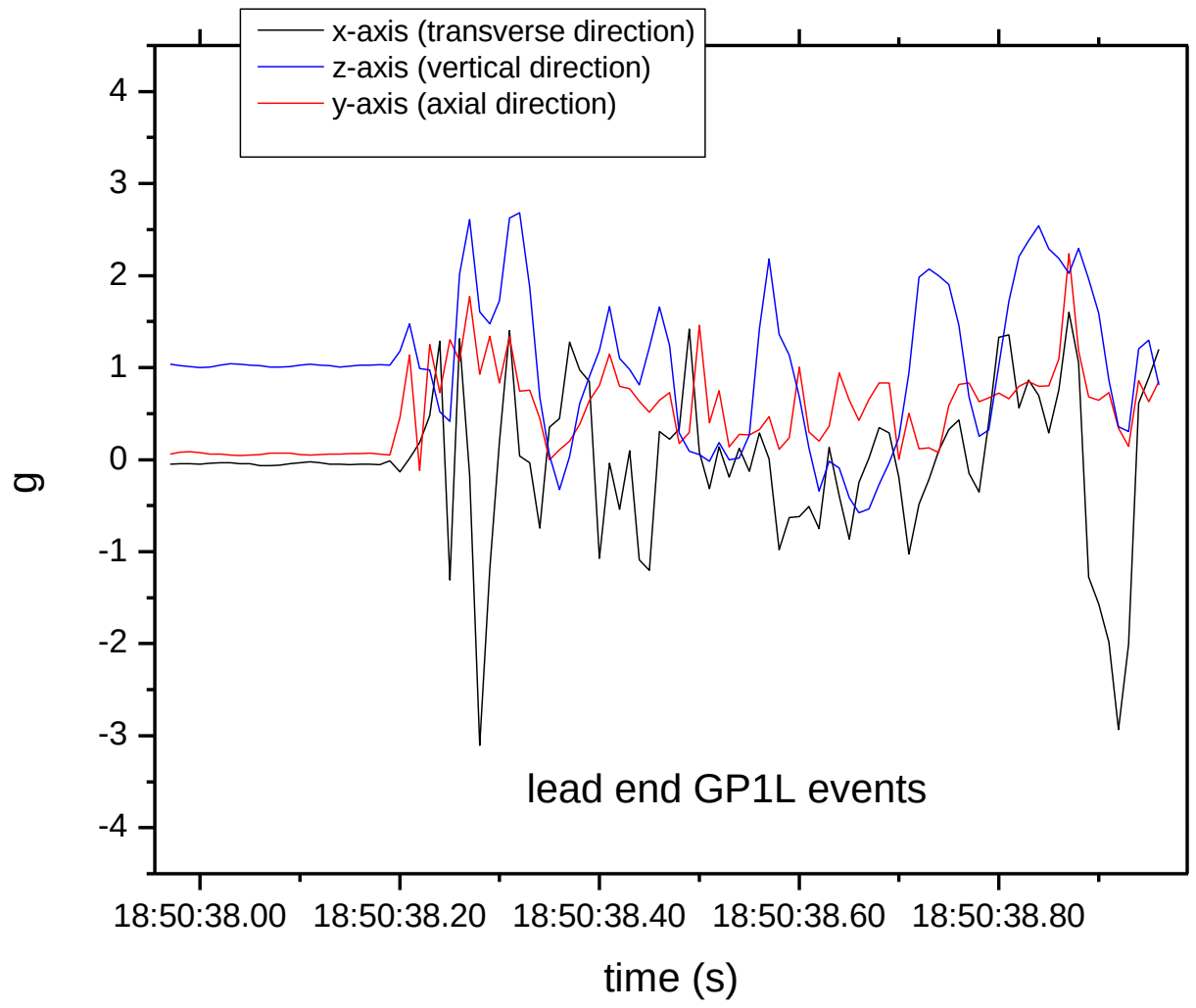

Figure 8: Two second event recorded by the lead end device. Event data are recorded when shocks are above 3 g.

## 6. Shock observed with Slamstick

The Slamstick device was installed on the non-lead end fixture clamp on the left side (see green circle in the magnet picture). This device has a piezoelectric and a Digital capacitive (DC-MEMS) accelerometers. Below there are the device settings:

Piezo: maximum shock threshold is 25g, sample rate is 1000 Hz

DC-MEMS maximum shock threshold is 16g, sample rate is 800 Hz

Each data set is 10 min long so the incident was captured in one data set. In Fig. 9 there is a snapshot of the shock vibrations collected during the accident with the DC-MEMS. The data are in good agreement with what is observed with the GP1L. Shocks above 3g cover a 1.5 s time window. The large oscillations observed afterwards are consistent with the incident dynamic and the detachment of the truck axle. The highest observed shock reaches 9 g along the vertical direction.

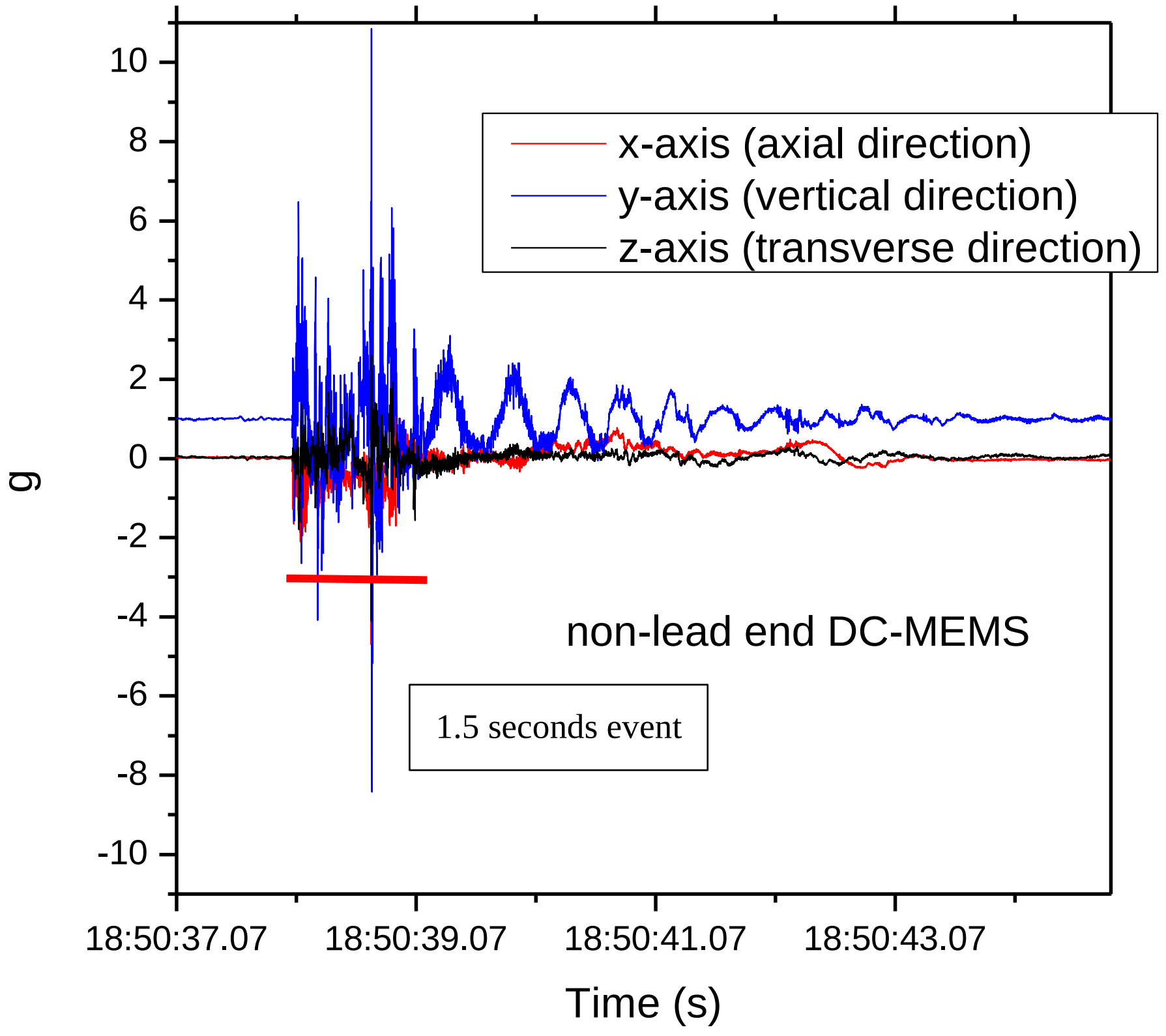

Figure 9: Shocks during the truck accident recorded by the DC-MEMS device installed on the non-lead end clamp

In Fig. 10 there is a zoom view of the highest recorded shocks. It is a high frequency oscillation that lasts around 15 ms. The highest peak (10g) is 2 ms long. The oscillation frequency is around 500Hz. This event could not be properly captured by the GP1L since its sample rate is around 10 ms.

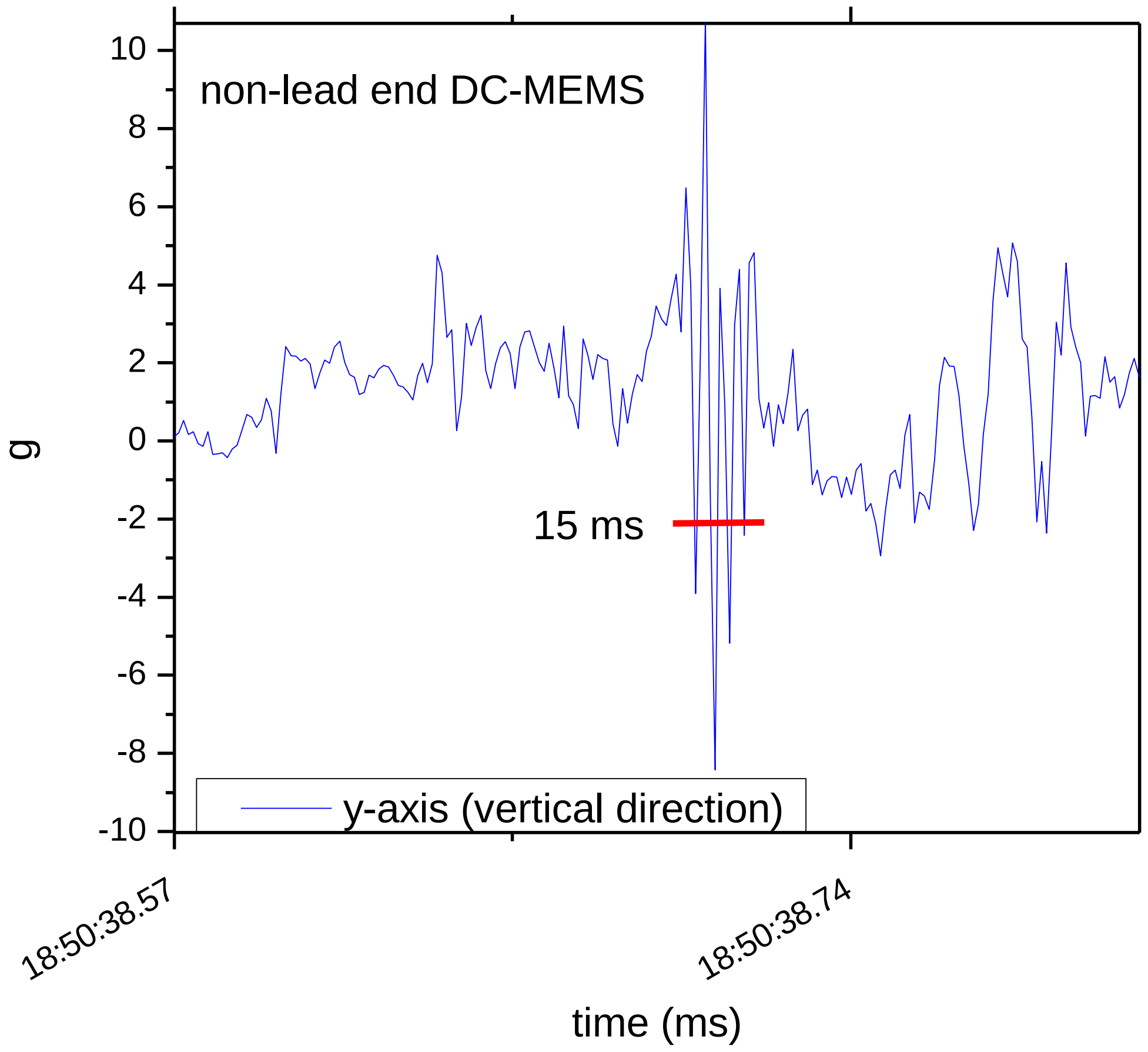

Figure 10: Zoom of the highest shocks region recorded during the truck accident by the DC-MEMS device installed on the non-lead end clamp

In the Fig.11, a shock data snapshot that covers 50 ms around the incident time is displayed for each DC-MEMS accelerometer direction. No shocks above requirements (5g) are observed along the transverse and axial direction.

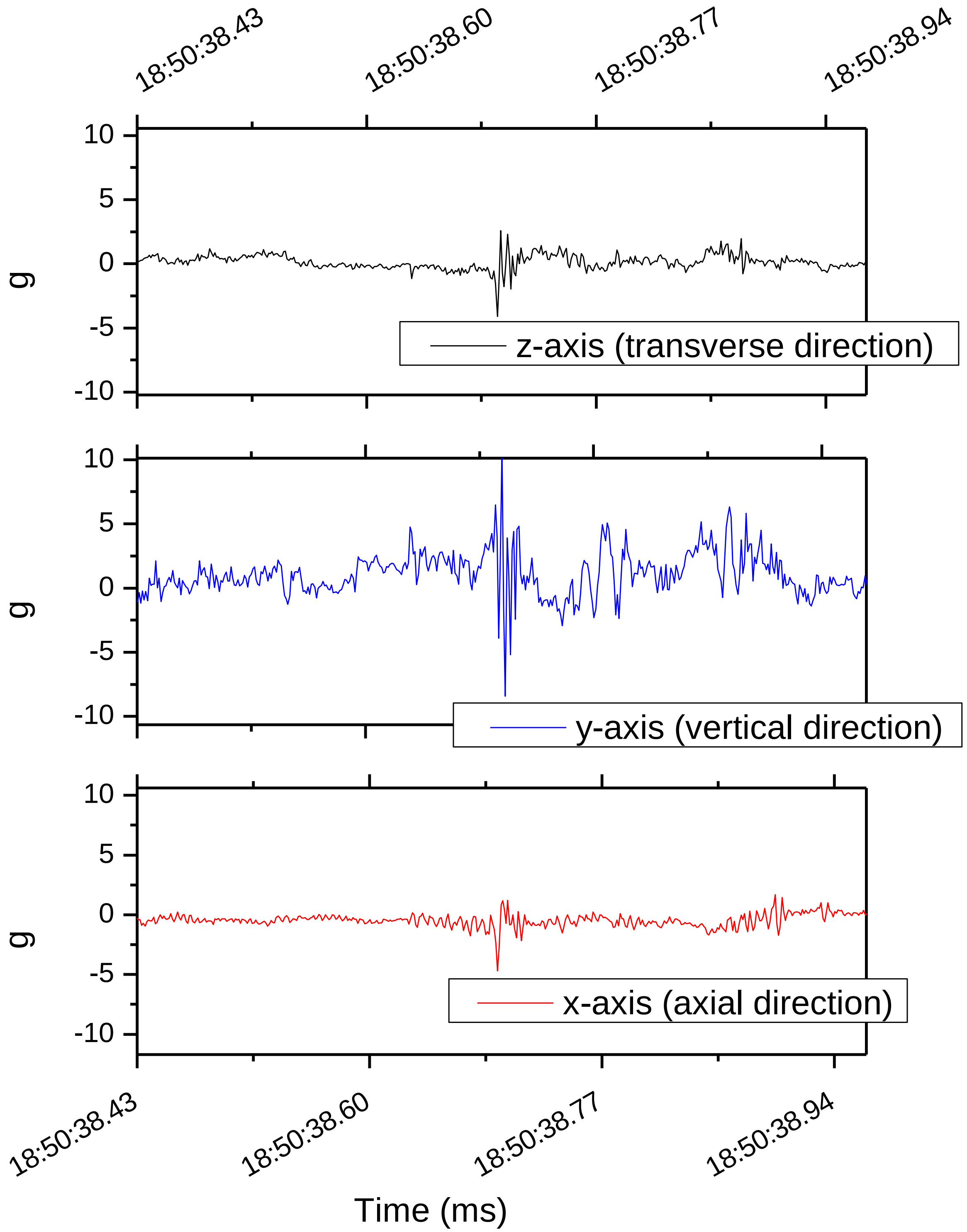

Figure 11: Shocks data for each direction recorded by the DC-MEMS. No shocks above requirements are observed over the axial and transverse direction

In Fig.12, the data collected with the piezo are shown. The highest shock is slight above 6 g in vertical direction. The difference between the Piezo and the DC-MEMS highest shocks value can be understood considering the actual bandwidth of those sensors. Sampling at 1000 Hz with the piezo, correspond to an antialiasing filter of 1/5$^{th}$ of the sample rate. The effective bandwidth is up to 200 Hz.

The DC-MEMS has a rough anti-aliasing filter, so data maybe be aliased. The anti-aliasing filter cutoff is ½ of the sample rate (500 Hz). The piezo-electric data are considered more reliable.

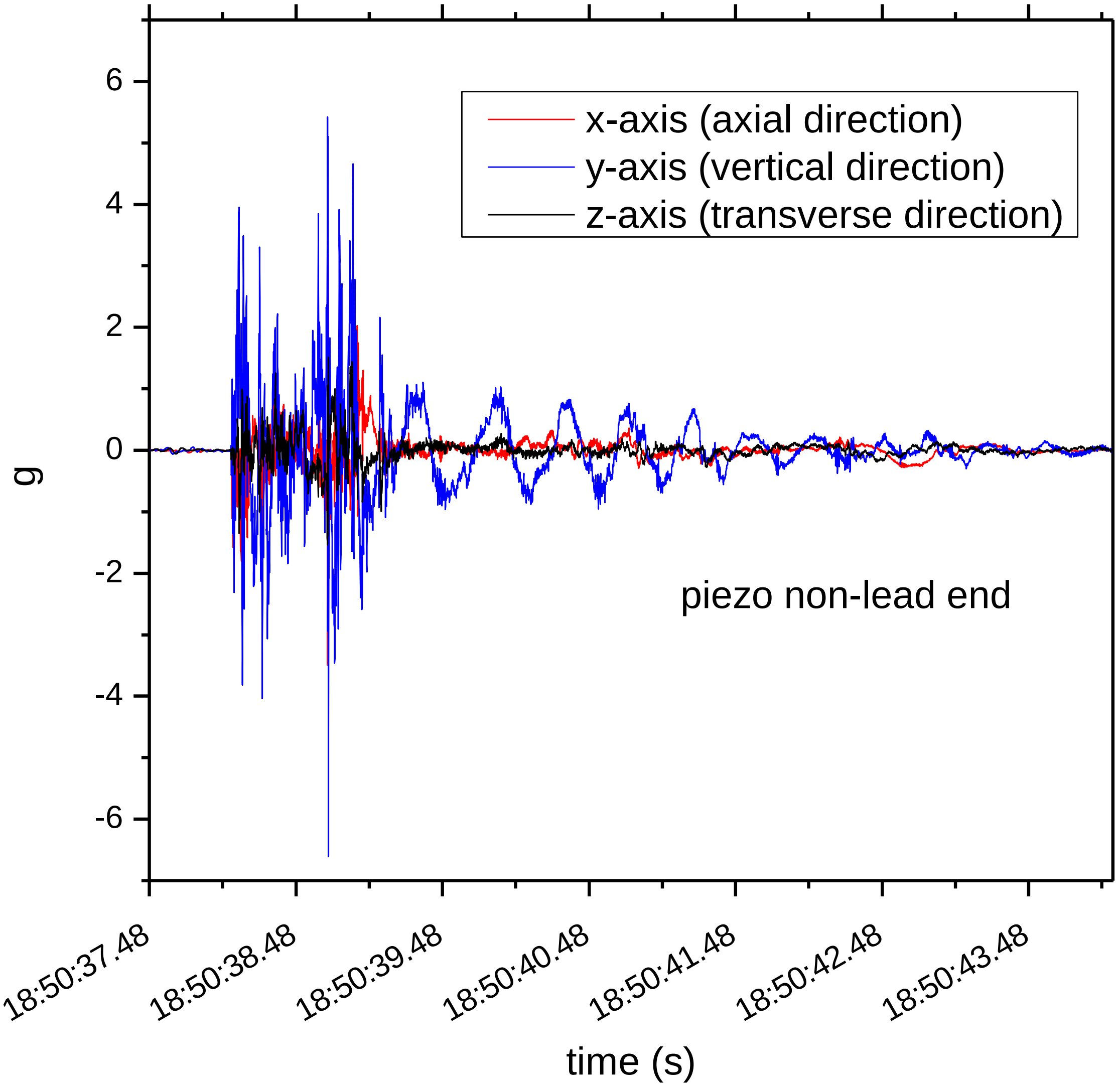

Figure 12: Shocks during the truck accident recorded by the Piezoelectric accelerometer device installed on the non-lead end clamp

In Fig. 13, a shock data snapshot of 5 seconds is displayed for each piezo accelerometer direction. No shocks above requirements (5g) are observed along the transverse and axial direction.

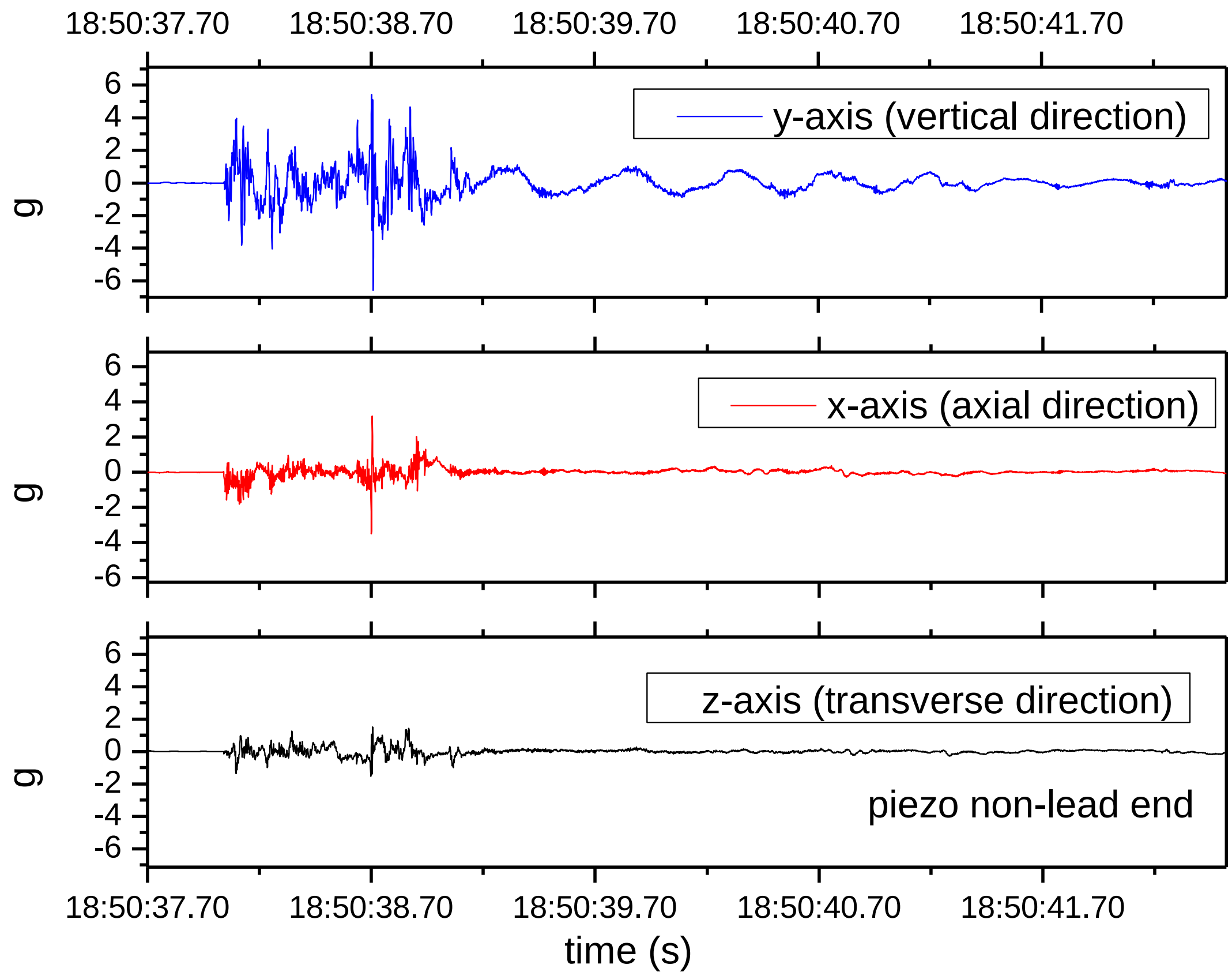

Figure 13: A five second snapshot of the shocks collected with the piezo devices. Shocks above requirement are observed only along the vertical direction

## 7. Fiber Bragg grating data

Two FBG fibers were installed on coil 222 (Q2) and 223 (Q1). Both fibers have 6 FBG glued on the coil pole in three positions: 700 (lead end), 1900 (center), 3800 (non-lead end). Axial and azimuthal strain are measured at each position. An extra FBG sensor for temperature compensation is present in the fiber of coil 222. Both fibers were broken before shipping and no further breaks were produced during the accident.

In Fig.14 and 15. a comparison between the FBG axial and azimuthal strain measured at LBNL and at FNAL is presented. The strain gauge data are also displayed. The strain data collected at FNAL are in good agreement with what have been observed before shipment. There is only the azimuthal strain value measure at position 1900 on coil 223 that displays a significant deviation.

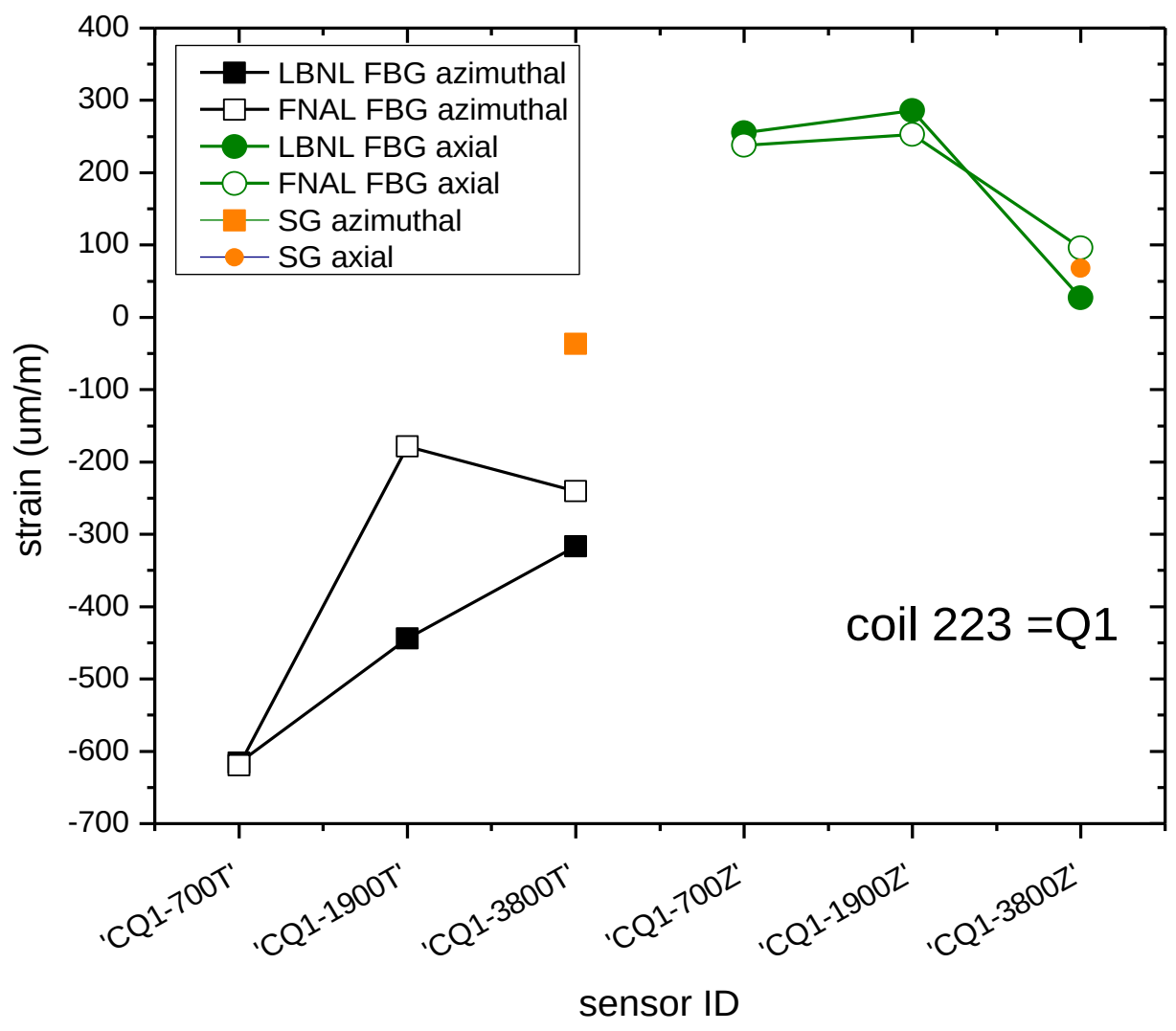

Figure 14: Comparison between azimuthal and axial strain data of coil 223 pole collected at LBNL and FNAL

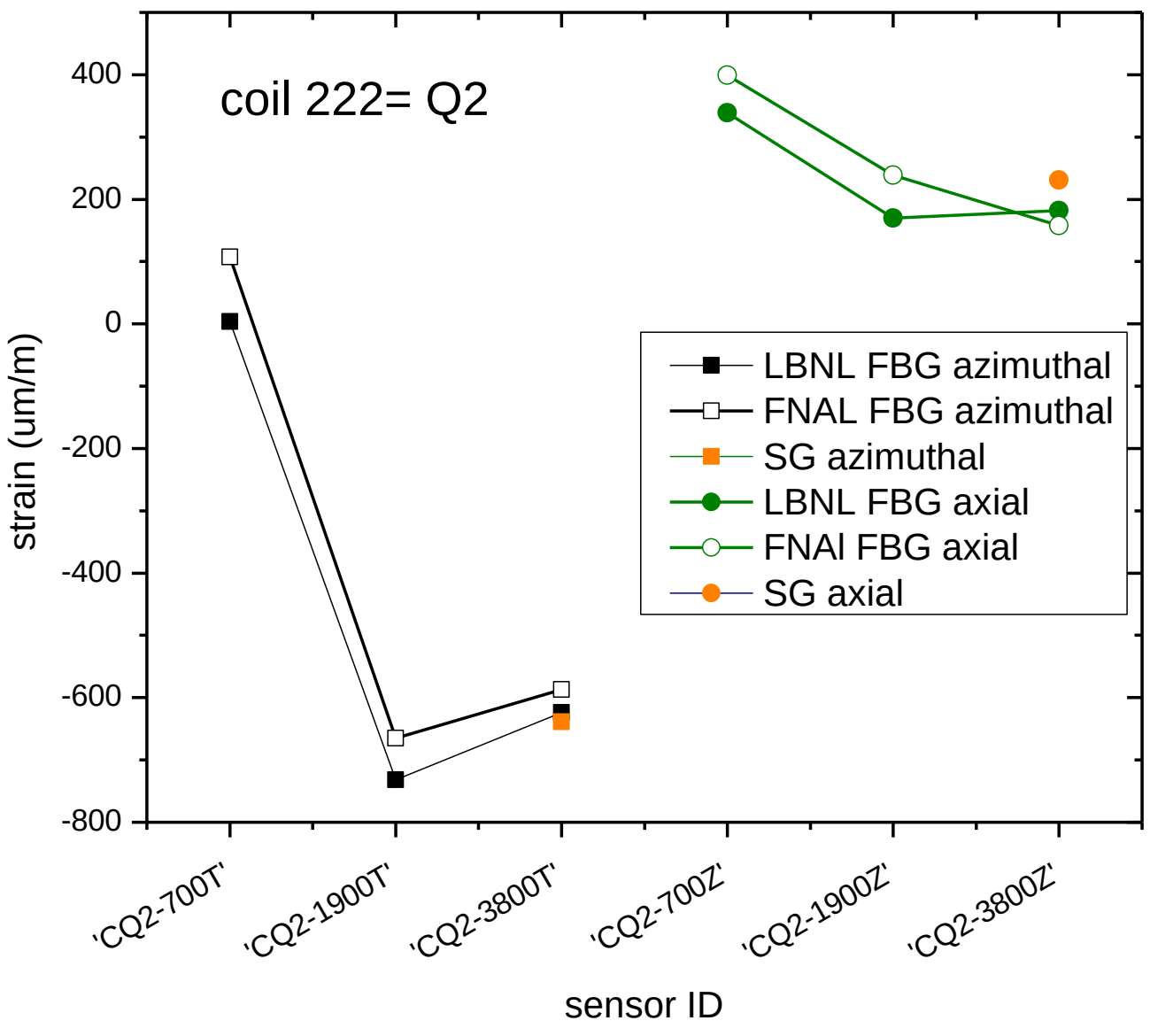

Figure 15: Comparison between azimuthal and axial strain data of coil 222 pole collected at LBNL and FNAL

## 8. Computational analysis

Conductor strain requirements during magnet handling and shipping are discussed in US-HiLumi-doc # 838). The conductor strain shall never exceed 500 microstrain.

The highest shocks observed by the magnet during the incident are around 6 g (piezo) and 10 g with the DC-MEMS. Those are dynamic events since the magnet sees 6g for 15 ms. A FEM analysis was performed in a static case using a 15g vertical load to estimate the amount of tensile strain supported by the magnet. This case is more conservative than the actual shock event (6 g for 15 ms).

A very small gap between yoke blocks starts to form at 15g as displayed in Fig. 16. The size of the gap is 0.0019 in (48 um).

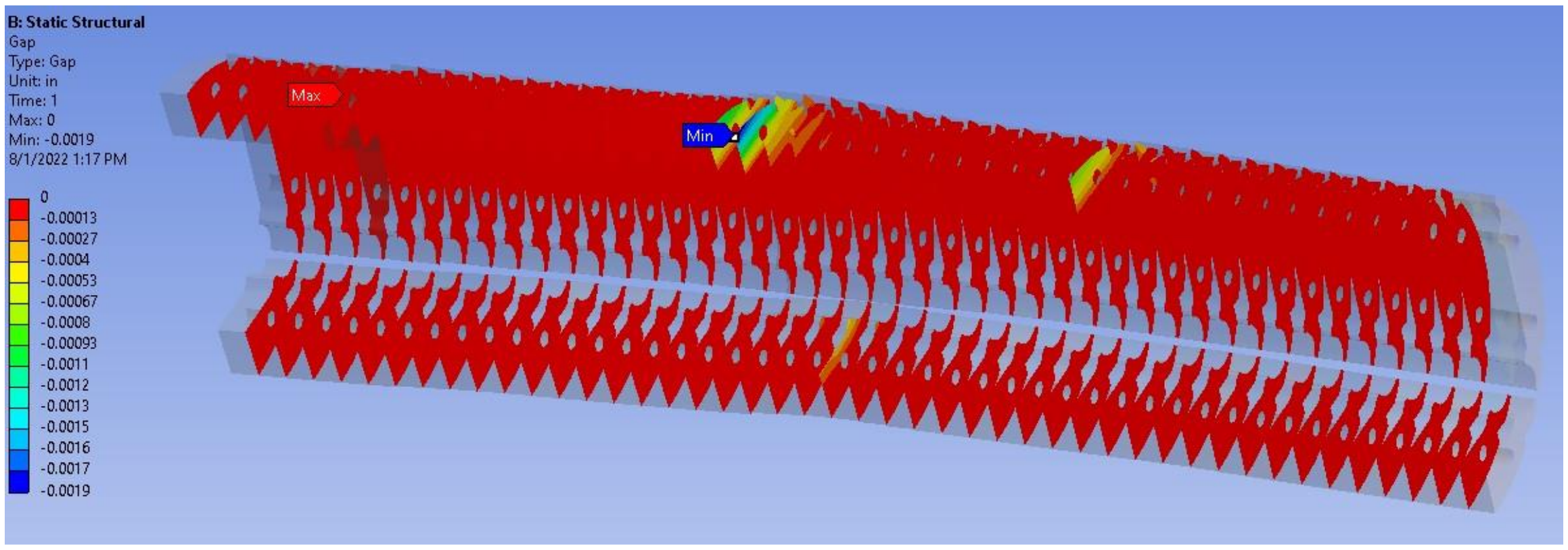

Figure 16: Yoke block gap forming with 15 g vertical load

The maximum coil tensile strain is found in that area and it is around 150 microstrain, which is well below the threshold requirements (500 microstrain).

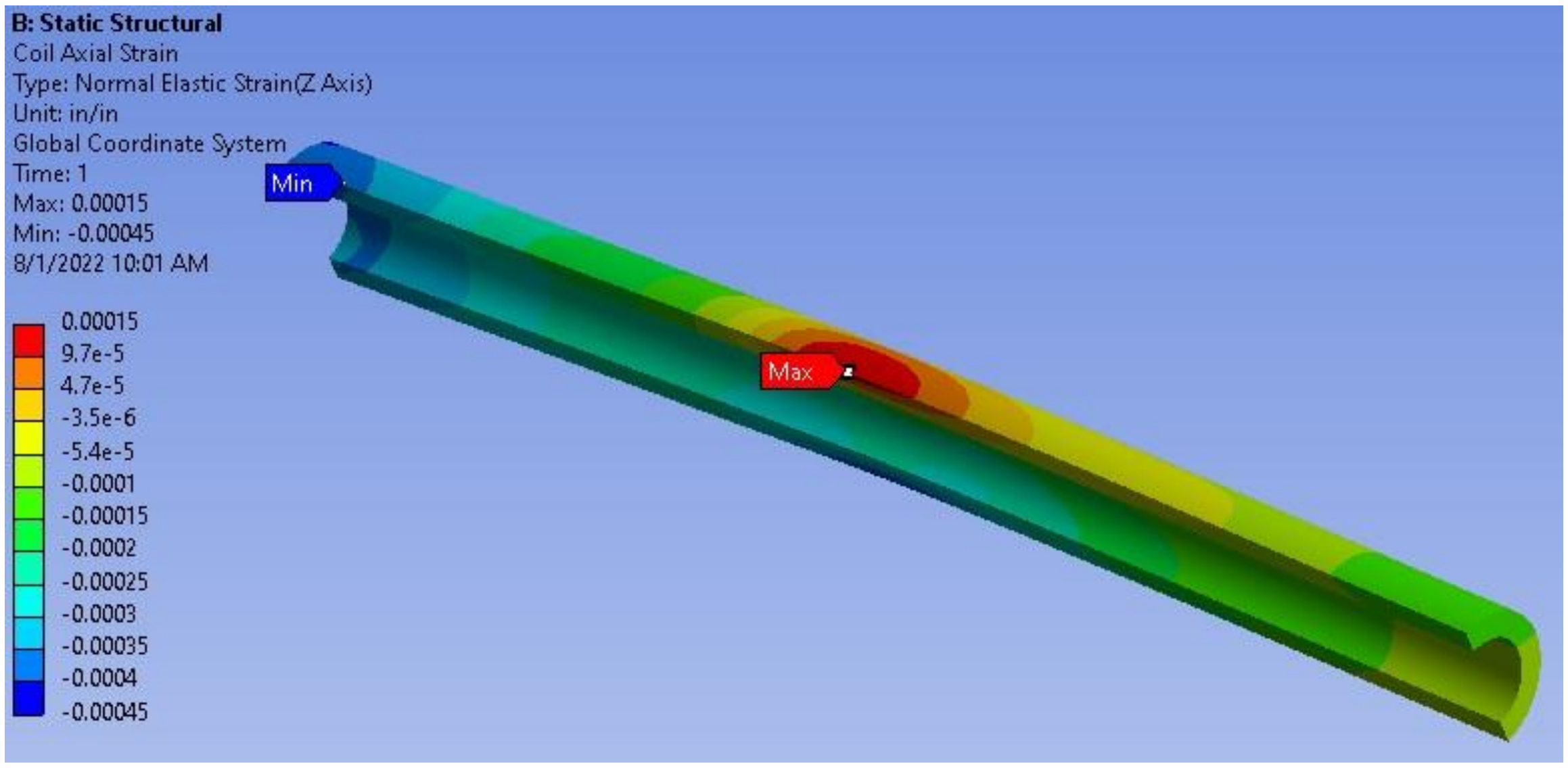

Figure 17: Tensile strain endured by the magnet structure

## 9. End plate slip verification against dynamic loads

The magnet longitudinal prestress is applied by stainless steel rods, bolted to two end plates. The end plate transfers the load to the magnet via bullets. A longitudinal prestress of 0.7 MN is applied at room temperature. The relative motion between the magnet and the end plate is prevented by friction. With a conservative friction coefficient of 0.1, we can assume that the end plate system will be able to sustain an in-plane (x-y) force of 70 kN before slipping with respect to the magnet. We can also assume, conservatively, that the whole magnet underwent a vertical acceleration of 15 g during the accident. The total weight of the longitudinal loading system is equal to 203 kg, which correspond to an inertia force of 30 kN. Comparing this to the frictional force available we can compute a safety factor as follows:

$$\eta = \frac{F_f}{F_i} = \frac{fF_z}{ma} = 2.34 \qquad \text{(a)}$$

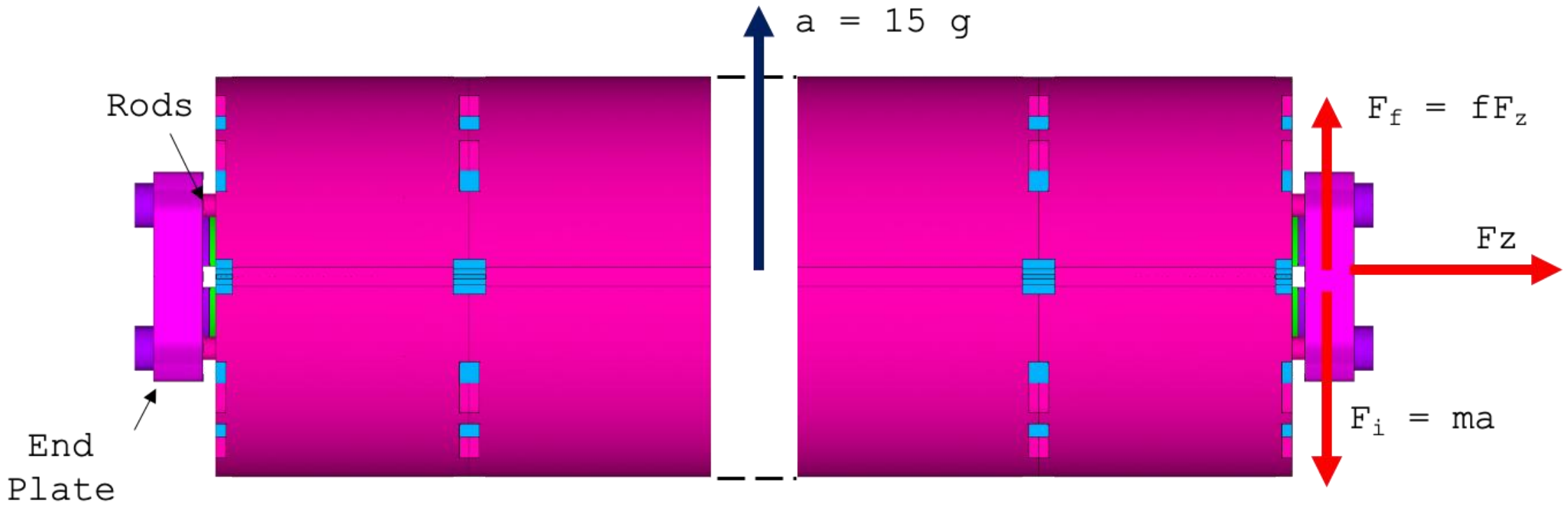

Figure 18 – Schematic of the magnet and end-plate system under the inertia forces due to the accident.

**B. First natural frequency of the shipping frame/magnet assembly**

A good approximation of the first natural frequency in the vertical direction can be computed knowing only the gravity sag $u_g$, and assimilating the assembly to a 1D spring-mass oscillator:

$$f_n = \frac{1}{2\pi}\sqrt{\frac{K}{m}} = \frac{1}{2\pi}\sqrt{\frac{g}{u_g}} \qquad \text{(b)}$$

where $g$ is the acceleration of gravity. Figure 11 shows the vertical deformation computed for a 15 g load, equal to 3.5 mm. From Eq. (x) we get a first natural frequency of 33 Hz.

## 10.Conclusions

The truck transporting MQXFA11 magnet was rear ended by another truck during the trip between LBNL and BNL The incident took place close to FNAL so the magnet was brought there for an investigation. This report summarizes the analysis work performed at FNAL to determine the condition of the magnet.

An initial visual inspection of the crate was performed at the tow yard. Both 5 and 10 g shock watches on the non-lead end passenger side of the trailer were tripped. The crate shifted and moved backward during the incident.

At FNAL a magnet visual inspection was performed, and no evident signs of damage were found. The magnet successfully passed the high voltage tests. Measurements of magnet straightness are in good agreement with those taken at LBNL before shipment. The same conclusion can be driven looking at the mechanical shim to axial restraint gap values.

Shocks above requirements were observed along the vertical direction on one of the three accelerometer devices installed on a MQXFA11 fixture clamp. The device was installed on the non-lead end clamp. The largest shock is in vertical direction. Its amplitude is ~6 g according to the piezo accelerometer, and 10 g according to the DC-MEMS accelerometer. This shock event is composed by high frequency oscillations (500 Hz) that last 15 ms. The duration of the highest shock is around 2 ms.

The difference between the Piezo and the DC-MEMS highest shocks value can be understood considering the actual bandwidth of those sensors and the anti-aliasing filter. The DC-MEMS has a rough antialiasing filter, so data maybe be aliased. The piezo-electric data are more reliable.

FBG fiber sensors data are in good agreement with what was measured before shipment.

FEM analysis was performed using a 15g static vertical load. The maximum tensile strain is around 150 microstrain which is below threshold requirements. The end plate slipping capacity was checked against friction and a safety margin of 2.34 was estimated.

The investigation performed at FNAL did not show any evident damage of the magnet. The computational analysis suggests that the conductor did not endure excessive strain. Considering those results, the MQXFA11 has been shipped to BNL for vertical testing.